\documentclass[10pt,journal=apchd5,manuscript=article,layout=onecolumn]{achemso}

\usepackage[version=3]{mhchem} 

\newcommand{\bb}[1]{\mathrm{#1}}

\author{Dmitry A. Svintsov}
\author{Aleksey V. Arsenin}
\author{Dmitry Yu. Fedyanin}
\email{dmitry.fedyanin@phystech.edu}
\affiliation[Unknown University]
{Laboratory of Nanooptics and Plasmonics, Moscow Institute of Physics and Technology, 9 Institutsky Lane, 141707 Dolgoprudny, Russian Federation}

\title[Full loss compensation in hybrid plasmonic waveguides under electrical pumping]
  {Full loss compensation in hybrid plasmonic waveguides under electrical pumping}

\abbreviations{SPP,MIS}
\keywords{Active plasmonics, surface plasmon amplification, metal-insulator-semiconductor structure, electrical pumping}

\begin{document}







\begin{abstract}
Surface plasmon polaritons (SPPs) give an opportunity to break the diffraction limit and design nanoscale optical components, however their practical implementation is hindered by high ohmic losses in a metal. Here, we propose a novel approach for efficient SPP amplification under electrical pumping in a deep-subwavelength metal-insulator-semiconductor waveguiding geometry and numerically demonstrate full compensation for the SPP propagation losses in the infrared at an exceptionally low pump current density of 0.8~kA/cm$^2$. This value is an order of magnitude lower than in the previous studies owing to the thin insulator layer between a metal and a semiconductor, which allows injection of minority carriers and blocks majority carriers reducing the leakage current to nearly zero. The presented results provide insight into lossless SPP guiding and development of future high dense nanophotonic and optoelectronic circuits.
\end{abstract}

\section{Introduction}

Surface plasmon polaritons (SPPs) being surface electromagnetic waves propagating along the interface between a metal and an insulator give a unique opportunity to overcome the conventional diffraction limit and thus can be used for light manipulation at the nanoscale \cite{Kriesch}. This gives plasmonic components a significant advantage over their photonic counterparts in the integration density and strength of light-matter interaction. However, high mode confinement is paid off by a significant localization of the SPP electromagnetic field in the metal and the resulting propagation length in nanoscale plasmonic waveguides does not exceed a few tens of micrometers due to absorption in the metal \cite{Kriesch,Conway,Sorger}. Nevertheless, it is possible to overcome this limitation by compensating ohmic losses with optical gain in the adjacent dielectric. Full loss compensation was successfully demonstrated by optical pumping \cite{Kena-Cohen,Oulton2009,Flynn}, which is easily implemented in a laboratory, but is impractical due to its very poor energy efficiency, stray illumination, and necessity of an external high power pump light source. In this regard, an efficient electrical pumping scheme is strongly needed for practical realization of SPP guides and circuits \cite{Fang2007,FedyaninAIP2010,FedyaninOE}.

In this paper, we propose a novel SPP amplification scheme based on minority carrier injection in metal-insulator-semiconductor (MIS) structures and demonstrate numerically full loss compensation in an electrically pumped active hybrid plasmonic waveguide. In contrast to the techniques based on heterostructures and quantum wells \cite{Costantini},  our approach gives a possibility to bring the active gain medium to the metal surface at a distance of a few nanometers. At the same time, a thin insulator layer can efficiently block the majority carrier current, which is quite high in metal-semiconductor Schottky-barrier-diode SPP amplifiers \cite{FedyaninNL}. This guarantees high mode confinement to the active region and very low threshold currents.

\section{Operating principle}

Realization of an electrical pumping scheme for loss compensation in SPP waveguides and cavities is a significant challenge, since there are at least two limitations in the development of plasmonic components. Firstly, plasmonics gives a possibility to reduce the mode size well below the diffraction limit, but, in order to use this opportunity, the component size should be comparable with the mode size. Accordingly, one has to use the SPP supporting metal-dielectric or metal-semiconductor interface as an electrical contact. Secondly, a limited choice of low-loss materials with negative real part of permittivity poses a serious challenge for efficient carrier injection, because gold, silver and copper typically form Schottky contacts to direct bandgap semiconductors. As a result, either the modal gain is rather low for full loss compensation, or the threshold current density is quite high~\cite{CostantiniThesis}, which leads to significant power consumption and low energy efficiency of the amplification scheme.

\begin{figure}[H]
\center{\includegraphics[width=0.8\linewidth]{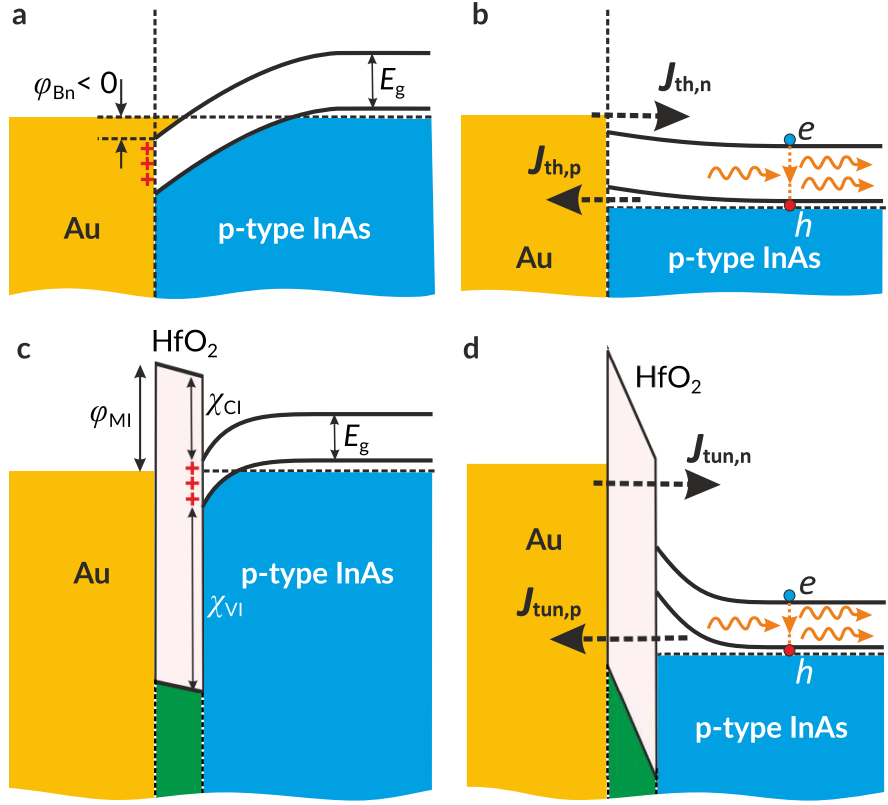} }
\caption{SPP amplification schemes based on a Schottky barrier diode \cite{FedyaninNL} and tunnel MIS contact. Top panels: Energy band diagrams for the Schottky contact between gold and indium arsenide in equilibrium (a) and at high forward bias (b). The Schottky barrier height for electrons $\varphi_{\bb{Bn}}$ at the Au/InAs contact is negative owing to the large density of surface states and, under high forward bias, electrons (minority carriers in p-type InAs) are freely injected in the bulk of the semiconductor. However, majority carriers (holes) also pass across the Au/InAs contact without any resistance, which results in a high leakage current. Bottom panels: Energy band diagrams for the tunnel Au/HfO$_2$/InAs MIS structure in equilibrium ($\mathrm{c}$) and at high forward bias (d). If the barrier height for holes $\chi_{\bb{VI}}$ is substantially greater than that for electrons ($\varphi_{\bb{MI}}$), the insulator layer can efficiently block majority carriers (holes), but be semi-transparent for minority carriers (electrons) escaping from the metal.}
\label{F1}
\end{figure}

The proposed technique based on an electrically pumped MIS structure (Figure~\ref{F1}) is remarkably different from the established approaches~\cite{Costantini, FedyaninOL, FedyaninNL, Hill2009}. First, SPP propagation losses are reduced by placing a very thin layer of a low refractive index insulator between the metal and semiconductor. In this case, the SPP electromagnetic field is pushed into the insulator reducing the portion of the SPP mode in the metal, which, however, does not impair the mode confinement~\cite{Sorger}. Second, and more important, a thin insulator layer can play a dual role: it is semi-transparent for electrons moving from the metal to semiconductor, which ensures favourable conditions for efficient minority carrier injection, and blocks the hole current. Such a unique feature gives a possibility to suppress high leakage currents \cite{Wade_Tunnel_injection_laser}, which are unavoidable in metal-semiconductor Schottky contacts \cite{FedyaninOL}, and not to form an ohmic contact to the semiconductor at the SPP supporting interface, as opposed to double-heterostructure and quantum-well SPP amplifiers \cite{CostantiniThesis}. At the same time, the proposed scheme provides high density of non-equilibrium carriers right near the metal surface, at the distance equal to the thickness of the insulator layer. At a high bias voltage, it becomes possible to create a sufficiently high electron concentration in the semiconductor and to satisfy the condition for population inversion \cite{CaseyPanish}, which provides gain for the plasmonic mode propagating in the MIS waveguide.

Previously, an attempt to describe the carrier transport in an electrically pumped hybrid plasmonic waveguide was made in Ref. \cite{Wijesinghe_wrong}. However, the authors assumed that both electrons and holes pass freely through the insulating layer and used the thermionic emission boundary conditions (literally the same as in the Schottky junction theory \cite{Green73,FedyaninOE}). This is erroneous in the presence of the insulating barrier and contradicts the well-established theory of current transfer through thin insulating films \cite{Stratton_basic_tunneling, Card_Rhoderick}. In fact, the insulating barrier strongly suppresses the thermionic current, while the main current component is due to tunneling.

The electron and hole tunnel injection currents $J_{\bb{t,cb}}$ and $J_{\bb{t,vb}}$ across the MIS contact can be calculated by integrating the flux of carriers incident on the contact timed by the transparency of the insulating barrier (see Methods). According to the energy band diagram shown in Figure \ref{F1} c,d, the ratio of the electron tunneling probability $D_{\bb{cb}}$ to the hole tunneling probability $D_{\bb{vb}}$ can be easily estimated as
\begin{equation}
\label{Dn_over_Dp_equation}
\frac{{\cal D}_{\bb{cb}}}{{\cal D}_{\bb{vb}}}=\exp \left[ 2^{3/2} \frac{m_{\bb i}^{1/2}{d_{\bb i}}}{\hbar } \left( \chi_{{\mathrm{VI}}}^{1/2}  - \varphi_{\mathrm{MI}}^{1/2} \right) \right],
\end{equation}
where $\chi_{\mathrm{VI}}$ is the valence band offset between the insulator and semiconductor, and $\varphi_{\mathrm{MI}}$ is the barrier height for electrons at the metal-insulator interface. As evident from this simple expression, in order to block only one type of carriers, the effective barrier height  for holes ($\chi_{\mathrm{VI}}$) must be substantially greater than that for electrons ($\varphi_{\mathrm{MI}}$) or vice versa. Hafnium oxide appears to be a very promising material for electron injection into the p-type III-V materials thanks to the low tunneling mass $m_{\bb i} = 0.1m_0$ \cite{Chang_HFO2_tunneling} and large valence band and small conduction band offsets. Both theoretical \cite{Robertson_band_offsets} and experimental \cite{Wheeler_Thesis} studies report $\chi_{\mathrm{CI}} = 2.4$~eV and $\chi_{\mathrm{VI}} = 3.2$~eV for the HfO$_2$/InAs interface. At the same time, $\varphi_{\mathrm{MI}} = 2.6$~eV for the gold contact \cite{Robertson_band_offsets}, and the ratio ${\cal D}_{\bb{cb}}/{\cal D}_{\bb{vb}}$ equals five, which is high enough for practical applications.

We next explore the carrier transport in the semiconductor. It is important to note that the electric potential is abruptly screened near the interface between the insulator and heavily doped semiconductor, which directly follows from the solution of Poisson's equation. At a doping density of $N_{\bb A} = 10^{18}$~cm$^{-3}$, the estimated screening length is equal to 2~nm, while the electron mean free path limited by impurity scattering equals $l_{\bb{fp}} = 10$~nm. This implies that electrons move ballistically through the screening region. Out of the screening region, the driving electric field is weak and the electron transport is governed by diffusion. Accordingly, the density distribution of the injected electrons is given by
\begin{equation}
\label{Diffusion_solution}
n ( z ) \approx \frac{J_{0,{\bb n}}}{e}\sqrt{\frac{\tau_{\bb R}}{D_{\bb n}}}\exp \left( -\frac{z}{L_{\bb D}}\right)+{n_0},
\end{equation}
where $J_{0,{\bb n}}$ is the current density at the insulator-semiconductor interface, $L_{\bb D}=\sqrt{\tau_{\bb R} D_{\bb n}}$ is the diffusion length, and $n_0$ is the equilibrium electron density. $L_{\bb D}$ is the most critical parameter determining the distribution of carrier density in InAs, and, consequently, the material gain profile. It depends on the electron mobility through the electron diffusion coefficient $D_{\bb n}$ and on the recombination rate through the electron lifetime $\tau_{\bb R}$. Equation (2) shows that, in order to compensate for the SPP propagation losses efficiently, $L_{\bb D}$ should be comparable with the half of the SPP penetration depth in InAs (500 nm at $\lambda =3.22$~$\mu$m) or greater than the latter. The electron mobility in p-type InAs is mainly limited by charged impurity scattering and is about  $\mu_{\bb n}=2\times 10^3$~cm$^2$V$^{-1}$s$^{-1}$ at 77~K at a doping level of $N_{\bb A}=10^{18}$~cm$^{-3}$~\cite{Sze_book}. The main contributions to the carrier recombination in InAs at 77 K come from the nonradiative Auger process $R_{\bb A} = (C_{\bb n} n + C_{\bb p} p)(np-n_0 p_0)$ and radiative recombination due to spontaneous emission $R_{\bb{sp}} = B_{\bb{sp}} (np-n_0 p_0)$, where $n_0$ and $p_0$ are the equilibrium electron and hole densities, $C_{\bb n} = C_{\bb p} = 10^{-27}$~cm$^6$/s are the Auger coefficients \cite{Abakumov_Perel_book}, and $B_{\bb{sp}}$ is the radiative recombination coefficient  calculated to be $4\times 10^{-10}$~cm$^3$s$^{-1}$. Since in a p-type semiconductor the density of electrons is much smaller than that of holes even at high injection levels, $\tau_{\bb R} =[C_{\bb p} N_{\bb A}^2 + B_{\bb{sp}} N_{\bb A}]^{-1} =700$ ps, which corresponds to a diffusion length of $L_{\bb D} = 950$ nm ensuring an excellent overlap between the distributions of the material gain (that is nearly proportional to the carrier concentration) and the SPP field.

The electron current $J_{0,{\bb n}}$ at the semiconductor-insulator junction is mainly determined by quantum mechanical tunneling through the insulating layer. In the case of an ideal interface, $J_{0,{\bb n}}$ is simply equal to the tunnel current $J_{\bb{t,cb}}$. However, the presence of intrinsic and extrinsic surface states at the semiconductor-insulator interface has a significant impact on the characteristics of the tunnel MIS contact (see Methods). First, surface states act as charge storage centers and affect the voltage drop across the insulator layer and band bending in the semiconductor. Second, these states are recombination centers \cite{Abakumov_Perel_book} and reduce the efficiency of electron injection. The electron current at the insulator-semiconductor interface $J_{0,{\bb n}}$ is smaller than the tunnel injection current $J_{\bb{t,cb}}$ by the value of the surface recombination current $J_{\bb{sr,n}}$. Third, direct tunneling from the metal to semiconductor through the surface states is also possible~\cite{Card_SSE, Lundstr?m_JAP}. This process contributes only to the majority carrier current and reduces the energy efficiency of the SPP amplification scheme. Fortunately, intrinsic surface states in InAs have a moderate density  $\rho_{\bb{ss}} \simeq 3\times 10^{12}$~cm$^{-2}$eV$^{-1}$~\cite{Noguchi_intrinsic_InAs}. Deposition of metal~\cite{Betti_MIGS_InAs} or insulator~\cite{Monch_surfaces} layers produces additional defects and  $\rho_{\bb{ss}}$ can be increased up to $10^{14}$~cm$^{-2}$eV$^{-1}$ at HfO$_2$/InAs interfaces~\cite{Wheeler_Dss}. Nevertheless, both intrinsic and extrinsic states can be treated by surface passivation before HfO$_2$ deposition. The reported values of $\rho_{\bb{ss}}$ at the trimethylaluminum-treated HfO$_2$/InAs interface is about $10^{13}$~cm$^{-2}$eV$^{-1}$~\cite{Wang_Dss_InAs}, while advanced oxygen-termination and surface reconstruction techniques give a possibility to reduce $\rho_{\bb{ss}}$ down to $2\times 10^{11}$~cm$^{-2}$eV$^{-1}$~\cite{Wang_APL}, which eliminates the influence of surface states on carrier transport in high quality samples.

\section{Active hybrid plasmonic waveguide in the passive regime}

\begin{figure}[H]
\center{\includegraphics[width=0.4\linewidth]{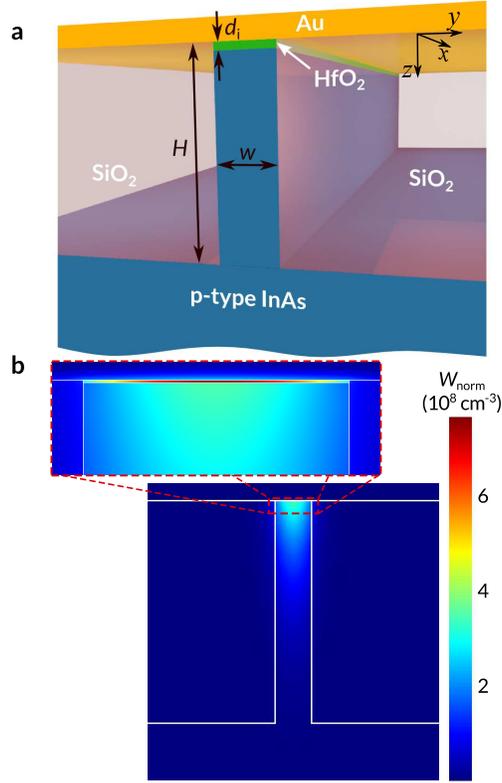} }
\caption{Schematic of a T-shaped hybrid plasmonic waveguide based on the Au/HfO$_2$/InAs MIS structure, $w$ is the waveguide width, $d_{\bb i}$ is the thickness of the low refractive index insulator layer between the metal and semiconductor, and $H$ is the waveguide height. (b) Distribution of the normalized energy density per unit length of the waveguide for the fundamental TM$_{00}$ mode at $\lambda$ = 3.22~$\mu$m, $H = 2.5$~$\mu$m, $w = 400$ nm and $d_{\bb i} = 3$ nm. The dielectric functions of the materials are as follows: $\varepsilon_{\mathrm{SiO_2}} = 2.00$ \cite{Malitson}, $\varepsilon_{\mathrm{HfO_2}} = 3.84$ \cite{Bright}, $\varepsilon_{\mathrm{InAs}} = 12.38$ \cite{Tsou} and  $\varepsilon_{\mathrm{Au}} = -545 + 38i$  \cite{FedyaninNL}.}
\label{F2}
\end{figure}

In order to achieve strong SPP mode localization and efficiently inject both electrons and holes into the active semiconductor region, a modified T-shaped plasmonic waveguide approach \cite{FedyaninNL} has been implemented [Figure \ref{F2} (a)]. InAs rib with an acceptor concentration of $10^{18}$~cm$^{-3}$ on InAs substrate is covered by a thin HfO$_2$ layer and surrounded by a low refractive index dielectric (SiO$_2$) to confine optical modes in the lateral direction. Finally, a metal layer is placed on top of the semiconductor-insulator structure to form an SPP supporting interface and a tunnel MIS contact for electron injection, while the substrate plays a role of the ohmic contact for majority carrier (hole) injection.
Two-dimensional eigenmode simulations using the finite element method reveal that the waveguide depicted in Figure \ref{F2}(a) supports a deep-subwavelength TM$_{00}$ SPP mode [Figure \ref{F2}(b)]. For the chosen waveguide dimensions, at a free space wavelength of 3.22~$\mu$m, which appears to be optimal for SPP amplification (see Section 4 for details), the effective index of the TM$_{00}$ mode is equal to 2.799 and the propagation length in the passive regime assuming the semiconductor to be lossless is about 66~$\mu$m corresponding to a modal loss of $2\mathrm{Im}\beta_{\bb{psv}} = 152$~cm$^{-1}$. The other modes supported by the T-shaped waveguide are very leaky photonic modes. In spite of poor field confinement to the lossy metal, propagation lengths of the photonic TM$_{10}$, TE$_{00}$, and TE$_{10}$ modes are 5.8~$\mu$m, 12.8~$\mu$m and 2.7~$\mu$m, respectively, which are much shorter than that of the TM$_{00}$ mode due to leakage into the high refractive index substrate. The rib height of $H = 2.5$~$\mu$m is close to the optimum, since, as $H$ decreases, the radiation loss of the plasmonic mode becomes substantially greater than the absorption in the metal and the SPP propagation length decreases down to 39~$\mu$m at $H = 2$~$\mu$m. On the other hand, greater rib heights do not provide a significant loss reduction, while high-aspect-ratio structures are difficult in fabrication. For the selected optimal geometrical parameters, the SPP mode demonstrates an exceptionally high level of confinement~\cite{LipsonGain} within the InAs region of more than 95\% provided by the small mode width and high group index.

\section{Full loss compensation in a deep-subwavelength waveguide}
Applying a negative voltage to the top Au electrode of the active plasmonic waveguide shown in Figure \ref{F2}a, one injects electrons into the p-type InAs region. This creates a high density of non-equilibrium electrons in the semiconductor rib, which reduces absorption in InAs \cite{CaseyPanish}. As the bias voltage increases, the quasi-Fermi level for electrons shifts upward towards the conduction band. When the energy difference between quasi-Fermi levels for electrons and holes exceeds the energy $\hbar \omega$ of the SPP quantum, InAs starts to compensate for the SPP propagation losses. The modal gain $G$ is given by the overlap integral of the material gain profile $g(y,z)$ and the electric field distribution of the SPP mode:
\begin{equation}
G = \frac{c \varepsilon_0 n_{\bb{InAs}} \int\limits_{-w/2}^{+w/2} {dy \int\limits_{d_{\bb i}}^H {dz g(y,z) |E(y,z)|^2 }}}{2 \int\limits_{-\infty}^{+\infty}{dy \int\limits_{-\infty}^{+\infty}{dz P_z(y,z)}}} - 2 {\mathrm{Im}}\beta_{\bb{psv}}.
\end{equation}
Here, $E(y,z)$ and $P_z(y,z)$ are the complex amplitudes of the electric field and the local power density of the SPP mode, respectively, and $2 {\mathrm{Im}}\beta_{\bb{psv}}$ is attributed to the SPP radiation and ohmic losses (see Section 3). In turn, the material gain coefficient $g(y,z)$ is related to the carrier quasi-Fermi levels $F_{\bb n}(y,z)$ and $F_{\bb p}(y,z)$ by the integral over transitions between the energy levels in the conduction and valence bands separated by the energy $\hbar \omega$ (see Methods). In heavily doped semiconductors, the material gain strongly depends on the impurity concentration \cite{Casey-Stern}, so are the electrical properties ($L_{\bb D}$, $\mu_{\bb n}$, and $\tau_{\bb R}$). The optimal acceptor concentration is found to be $10^{18}$~cm$^{-3}$. As the doping concentration increases, the Auger recombination rate and the impurity scattering rate rapidly increase greatly reducing the electron diffusion length. On the other hand, at lower doping levels, the number of free electron states in the valence band is insufficient to provide a pronounced gain upon electrical injection. At the given impurity concentration and a reasonably high density of injected electrons of $2\times 10^{16}$~cm$^{-3}$ (which corresponds to $F_{\bb n} - E_{\bb c} = k_{\bb B} T$), the material gain spectrum exhibits a maximum of 310~cm$^{-1}$ at $\hbar \omega =0.385$~eV ($\lambda =3.22$~$\mu$m), sufficiently high for the net SPP amplification.

Figure 3 shows the simulated gain-current characteristic of the plasmonic TM$_{00}$ mode in the active hybrid plasmonic waveguide for the HfO$_2$/InAs interfaces of different quality. All curves exhibit the same trend. At zero bias, the concentration of holes in the semiconductor is many orders of magnitude greater than that of electrons, InAs strongly absorbs the SPP propagating in the waveguide and the modal loss is two-and-a-half times greater than in the case of the lossless (e.g. wide bandgap) semiconductor. As the bias voltage increases, electrons are injected in the InAs rib (Figure 4a), which reduces absorption in the semiconductor near the MIS contact, but still, InAs strongly absorbs at a distance greater than $L_{\bb D}/2$ (Figure 4b). Nevertheless, the SPP modal loss steadily decreases with the injection current (this region is shown in blue in Figure 3). At some point (green curve in Figure 4b), the overlap between the distributions of the material gain and the SPP field is zero and, at higher bias voltages, the gain in InAs produced by injected electrons partially compensates for the SPP propagation losses. 
\begin{figure}[H]
\center{\includegraphics[width=0.5\linewidth]{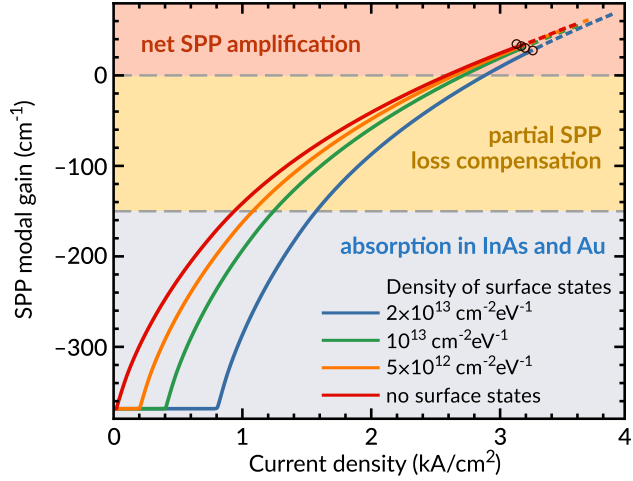} }
\caption{
SPP modal  gain versus pump current for different densities of surface states at the HfO$_2$/InAs interface. Blue, yellow and red regions show three regimes of the active plasmonic waveguide: at low injection currents, the electron concentration in InAs is quite small for population inversion; in the yellow region, the SPP propagation losses are partially compensated by optical gain in InAs; and at high injection currents, the material gain is significantly large for net SPP gain. The black circles highlight the breakdown condition for the HfO$_2$ insulating layer  $V_{\bb i}/d_{\bb i}=1$ V/nm \cite{Chang_HfO2_breakdown}.}
\label{F3}
\end{figure}
Eventually, at an injection current of about 2.7~kA/cm$^2$ (Figure 3), ohmic losses and radiation losses of the plasmonic mode are fully compensated. For comparison, previously, net SPP amplification was reported at current densities of the order of 20~kA/cm$^2$ in waveguide structures based on the Au/p-InAs Schottky contact~\cite{FedyaninNL}. Here, we have achieved one order of magnitude reduction in the threshold current density using the tunneling Au/HfO$_2$/InAs contact instead of the Schottky contact and blocking the majority carrier current with a thin insulator layer.

As evident from Figure 3, the current density required for full loss compensation is robust against the quality of the HfO$_2$/InAs interface. It is equal to 2.6 kA/cm$_2$ in the case of the ideal interface and increases by only 12\% as the density of surface states increases to $2\times 10^{13}$~cm$^{-2}$eV$^{-1}$. This small increase is completely attributed to the tunnel current from the metal to surface states (Figure 4a), while the surface recombination current $J_{\bb{sr,n}}$ is more than six orders of magnitude lower than the total current. This is explained by the energy barrier for electrons in the semiconductor due to band bending at the HfO$_2$/InAs interface under high forward bias (see inset in Figure 4b)~\cite{Rimmer_SR_reduction}.

\begin{figure}[H]
\center{\includegraphics[width=0.5\linewidth]{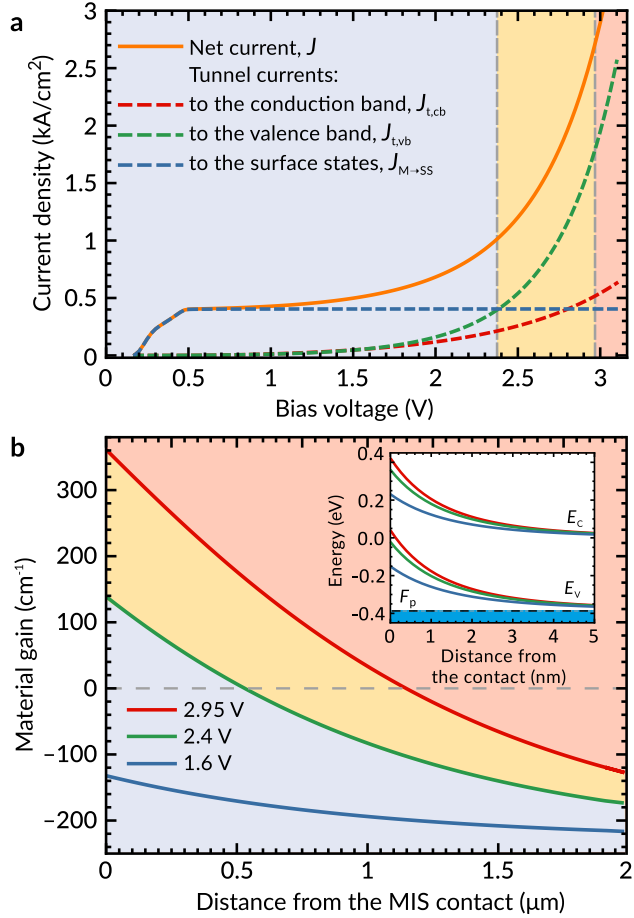} }
\caption{(a) Contributions of different tunneling processes to the injection current. (b) Material gain profile across the InAs rib at different bias voltages. Inset: simulated energy band diagrams near the HfO$_2$/InAs interface. For both panels, the density of surface states equals $10^{13}$~cm$^{-2}$eV$^{-1}$.}
\label{F4}
\end{figure} 

Interesting is the possibility to control optical and electrical properties of the active T-shaped hybrid plasmonic waveguide by varying the thickness of the insulator layer. As $d_{\bb i}$ increases, the portion of the SPP field in the metal decreases, which results in a decrease of ohmic losses. At the same time, thicker insulating layers block the majority carrier current more efficiently as follows from Equation~(1). However, it appears that the tunnel current is very sensitive to the thickness of the HfO$_2$ layer and $d_{\bb i} = 3$ nm is the optimal value. As $d_{\bb i} $ is decreased by 0.5 nm (one atomic layer), the majority carrier current drastically increases due to direct tunneling from the metal to surface states and, at a surface-state density of $10^{13}$~cm$^{-2}$eV$^{-1}$, the SPP propagation losses are fully compensated at a very high current density of 7.5~kA/cm$^2$ 
. On the other hand, the 3.5 nm thick HfO$_2$ layer efficiently blocks the majority current, but the barrier width for minority carriers is also increased and higher bias voltage are needed for efficient electron injection. As a result, the electric field in HfO$_2$ rapidly reaches the breakdown value of 1 V/nm \cite{Chang_HfO2_breakdown} as the electron current increases. The minimum achievable SPP modal loss is 240~cm$^{-1}$ that is much greater than  $2\mathrm{Im}\beta_{\bb{psv}} = 155$~cm$^{-1}$ evaluated in the case of the lossless (e.g. wide bandgap) semiconductor. For the optimal HfO$_2$ layer thickness of 3 nm, the breakdown condition $V_{\bb i} = 3$ V corresponds to the net SPP modal gain of 40~cm$^{-1}$, which can be increased up to 160~cm$^{-1}$ in high-quality samples, where the breakdown electric field can be as high as 1.2 V/nm~\cite{Chang_HfO2_breakdown}. 


\section{Low-current loss compensation in copper waveguide}

The effective density of states in the valence band of InAs is 75 times greater than in the conduction band. Therefore, in spite of the relatively large ratio ${\mathcal{D}}_{\bb{cb}}/{\mathcal{D}}_{\bb{vb}}=5$ for the Au/HfO$_2$/InAs contact, the electron tunnel current does not exceed that of holes (Figure 4a). In addition, under high forward bias, the effective barrier height for holes, which can be estimated as the energy difference between the hole quasi-Fermi level in InAs near the insulator-semiconductor contact and the valence band edge of HfO$_2$ at $z = d_{\bb i}$, is substantially reduced due to band bending (inset in Figure 4b) and the ratio $J_{\bb{t,cb}}/J_{\bb{t,vb}}$ rapidly decreases as the bias voltage increases (Figure 4a). To improve the efficiency of the amplification scheme, one should increases the difference between barrier heights $\chi_{\bb{VI}}$ and $\varphi_{\bb{MI}}$ as it can be seen from equation~(1). To satisfy this demand, one or more materials in the T-shaped active plasmonic waveguide (Figure 2a) should be replaced.

It would be naive to expect better characteristics in plasmonic structures based on metals different from gold and silver, however, in electrically pumped active plasmonic waveguides, both electrical and optical processes are involved, and the optimal configuration is typically the result of the interplay between them. Copper is characterized by higher absorption at optical frequencies than gold~\cite{Ordal, Lenham} and the modal loss of the plasmonic TM$_{00}$ mode in the T-shaped waveguide increases from  152~cm$^{-1}$ to 189~cm$^{-1}$. On the other hand, the electron work function of copper is equal to 4.7~eV, which is 0.4~eV lower than that of gold. Such a small difference is sufficient to produce an appreciable effect on carrier tunneling through the insulating barrier and increase the efficiency of minority carrier injection in InAs. The electron-to-hole transparency ratio ${\mathcal{D}}_{\bb{cb}}/{\mathcal{D}}_{\bb{vb}}$ of the Au/HfO$_2$/InAs contact is as large as 20, and the electron tunnel current $J_{\bb{t,cb}}$ substantially exceeds the hole tunnel current $J_{\bb{t,vb}}$ (Figure 5). 

At a surface-state density of $5\times 10^{12}$~cm$^{-2}$eV$^{-1}$, the propagation losses of the plasmonic mode are fully compensated at a current density of 0.95~kA/cm$^2$, which is significantly smaller than in the case of the gold active plasmonic waveguide despite higher absorption losses of the SPP mode. Passivation of the InAs surface reduces the density of surface states by orders of magnitude~\cite{Wang_APL}, and the regime of full loss compensation is reached at a very low current $J = 0.82$~kA/cm$^2$, which is comparable with the threshold currents in double-heterostructure InAs lasers~\cite{Aydaraliev}. In addition, the copper waveguide operates at smaller bias voltages than the gold one. First, this results in lower energy consumption per unit waveguide length, which decreases from 30 mW/mm to 8.2~mW/mm in the regime of full loss compensation. Second, and more importantly, smaller bias voltages create lower electric fields in the HfO$_2$ layer. The breakdown condition ($V_{\bb i} / d_{\bb i} = 1$ V/nm) corresponds to the net SPP modal gain of 220~cm$^{-1}$ that is much higher than in the similar gold waveguide. This allows one to use the proposed amplification scheme in nanolasers \cite{Stockman}, where not only ohmic, but also high radiation losses have to be compensated.

\begin{figure}[H]
\center{\includegraphics[width=0.6\linewidth]{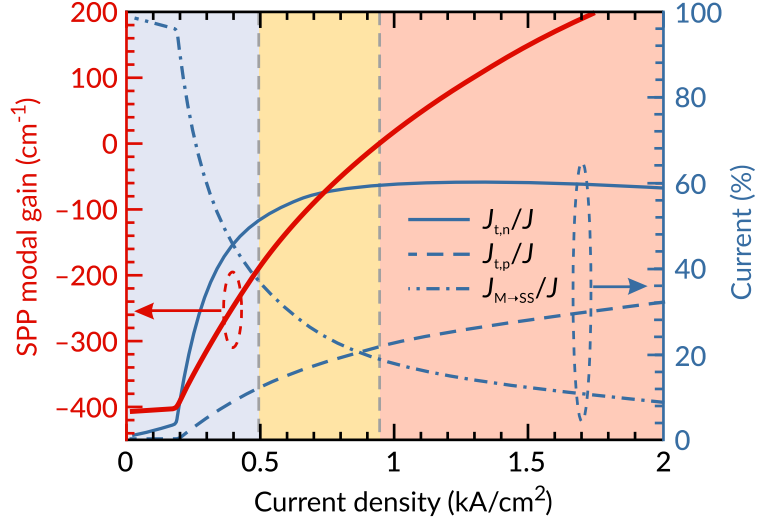} }
\caption{Gain-current characteristic for the SPP mode of the T-shaped active hybrid plasmonic waveguide based on the Cu/HfO$_2$/InAs MIS structure and relative contributions of different tunnelling processes to the total current. The density of surface states is equal to $5\times 10^{12}$~cm$^{-2}$eV$^{-1}$ and the dimensions of the waveguide are the same as in Figure 2b.}
\label{F5}
\end{figure} 

\section{Conclusion}
In summary, we have proposed a novel approach for SPP amplification in deep-subwavelength structures under electrical pumping and demonstrated full loss compensation in an active hybrid plasmonic waveguide. The amplification scheme is based on efficient minority carrier injection in metal-insulator-semiconductor contacts at large forward bias, which creates a population inversion in the semiconductor and provides optical gain for the plasmonic mode propagating along the metal interface. At the same time, a thin insulator layer blocks majority carriers giving a possibility to reduce the leakage current to that in double-heterostructure lasers. Comprehensive electronic/photonic simulations show that the SPP propagation losses can be fully compensated in the electrically driven active hybrid plasmonic waveguide based on the Au/HfO$_2$/InAs structure at a current density of only 2.6~kA/cm$^2$. Moreover, by replacing gold with copper, one significantly improves the efficiency of minority carrier injection, and, in spite of higher ohmic losses, the current density in the regime of lossless SPP propagation is reduced down to 0.8~kA/cm$^2$. Such an exceptionally low value demonstrates the potential of electrically pumped active plasmonic waveguides and plasmonic nanolasers for future high-density photonic integrated circuits.

\section{Methods}
{\textbf{Evaluation of the tunnel current}}. Tunnel current density across the MIS contact is given by~\cite{Stratton_basic_tunneling}
\begin{equation}
\label{Tunnel_injection_equation}
{J_{\bb{t},k}}=\frac{4 \pi e}{(2 \pi \hbar)^3}\int\limits_{E_{\bb c}}^{+\infty }{dE\left[ {f_{\bb m}}(E) - f_k(E) \right]}\int\limits_0^{p_{\bot ,\max}}{p_{\bot } dp_{\bot }{{\mathcal D}_k}\left( E,{p_{\bot }} \right)}.
\end{equation}
Here, the subscript $k$ indicates either the conduction (cb) or valence (vb) band, $f_{\bb m}(E)$, $f_{\bb{cb}}(E)$, and $f_{\bb{vb}}(E)$ are the Fermi-Dirac electron distribution functions in the metal and respective semiconductor bands, $p_{\bot\max}$ is the maximum transverse momentum of the electron in the semiconductor at a given energy $E$, ${\cal D}_k (E,p_{\bot})$ is the probability of carrier tunneling to the $k$-th band, $E_{\bb c}$ is the conduction band edge in the semiconductor, $\hbar$ is the Planck's constant and $e$ is the elementary charge.

The tunneling probability ${\cal D}_k (E,p_{\bot})$ is obtained in the WKB-approximation
\begin{equation}
\label{DEp_equation}
{\cal D}_k \left( E,p_{\bot } \right)= \exp \left\{ -\frac{2}{\hbar} \operatorname{Im}
\int\limits_{0}^{d_{\bb{i}}} {\sqrt{2 m_{\bb{i}} \left[ E-\frac{p_{\bot}^2} {2m_{\bb i}}-U_k (z) \right]}dz} \right\},
\end{equation}
where $U_k(z)$ is the conduction band edge [$U_{\bb{cb}}(z)$] or the valence band edge [$U_{\bb{vb}}(z)$] of the insulator, $m_{\bb i}$ is the effective electron mass in the insulator (tunneling mass), and $d_{\bb i}$ is its thickness. 

{\textbf{Accounting for the surface states at the HfO$_2$/InAs interface.}} The surface charge and spurious currents associated with surface states can be found from the detailed balance between three major processes 

(1) electron capture from the conduction band of the semiconductor to the surface states and their thermal emission to the band;

(2) hole capture and thermal emission;

(3) direct tunnelling from the metal to surface states.

Following Landsberg and Klimpke \cite{Landsberg_surf_recomb}, we obtain the occupancy $f_{\bb{ss}}$ of the surface state with energy $E$ in a steady state:
\begin{equation}
\label{SS_Occupation}
f_{\bb{ss}} ( E )=\frac{\nu_{\bb{n}} + \nu_{\bb{p}} {e^{\left( F_{\bb p}-E \right)/k_{\bb{B}}T}} + \nu_{\bb{t}} (E) f_{\bb{m}} (E)} {\nu_{\bb{n}}\left[ e^{( E-F_{\bb{n}})/k_{\bb{B}} T} + 1 \right]+\nu_{\bb{p}} \left[ {e^{( {F_{\bb{p}}}-E)/k_{\bb{B}}T}}+1 \right]+\nu_{\bb{t}}(E)}.
\end{equation}

In the above expression, we have introduced the rates $\nu_{\bb n} = n|_{z=d_{\bb i}}\sigma_{\bb n} v_{\bb n}/4$, $\nu_{\bb p} = p|_{z=d_{\bb i}}\sigma_{\bb p} v_{\bb p} / 4$, and $\nu_{\bb t} (E)$ characterizing the abovementioned population and depopulation processes; $\sigma_{\bb n}$ and $\sigma_{\bb p}$ are the electron and hole capture cross sections, $v_{\bb n}$ and $v_{\bb p}$ are the electron and hole thermal velocities, and $\nu_t$ is the rate of direct tunnel transitions from the metal to surface states 
. The latter can be expressed as $\nu_t (E)=\tau_0^{-1}{\cal D}(E,0)$ \cite{Lundstr?m_JAP}, where ${\cal D}(E,0)$ is the barrier transparency evaluated at zero transverse momentum, and $\tau_0 \simeq 0.25$ fs is the characteristic electron escape time from the surface state to the metal. Experimental studies reveal the donor-type behavior of surface defects in InAs~\cite{Monch_surfaces}. In the neutral state, their hole capture cross-section $\sigma_{\bb p}$ is about $10^{-15}$ cm$^{-2}$~\cite{Abakumov_Perel_book}, while, in the charged state, the electron capture cross-section sn is estimated to be $10^{-16}$ cm$^{-2}$ using the theory of electron capture mediated by cascade phonon emission~\cite{Abakumov_JETP}. Such a small cross-section is attributed to the slow phonon emission rate, which is inseparably linked with the high electron mobility of $10^{5}$~cm$^2$V$^{-1}$s$^{-1}$ in intrinsic InAs \cite{Adachi_book}. 

Assuming the density of surface states $\rho_{\bb{ss}}$ per unit area per unit energy to be uniform over the band gap, one can write expressions for the surface charge density
\begin{equation}
\label{Qss}
{Q_{\bb{ss}}}=e{\rho_{\bb{ss}}}\int\limits_{E_{\bb c}}^{E_{\bb v}}{dE\left[ 1 - f_{\bb{ss}} (E) \right]},
\end{equation}
the current density associated with surface recombination
\begin{equation}
\label{Jsr}
{J_{\bb{sr,n}}}=\frac{1}{4}ev_{\bb n} \left.n\right|_{z=d_{\bb i}}{\sigma_{\bb n}}{\rho_{\bb{ss}}}
\int\limits_{E_{\bb c}}^{E_{\bb v}}{dE}\left\{ \left[ 1 - f_{\bb{ss}}(E) \right]-{f_{\bb{ss}}}(E){{e}^{\left( E-F_{\bb n} \right)/k_{\bb B}T}} \right\},
\end{equation}
and the direct tunneling current from the metal to the surface states
\begin{equation}
\label{Jmss}
{J_{\bb{t,ss}}}=e{\rho_{\bb{ss}}}\int\limits_{E_{\bb c}}^{E_{\bb v}}{dE \nu_{\bb t}(E) \left[ {f_{\bb m}m}(E) - f_{\bb{ss}}(E) \right]},
\end{equation}
Finally, equations (2), (4), (7), (8), and (9) are solved with Poisson's equation 
.

{\textbf{Calculation of the material gain}}. Optical gain in the semiconductor is obtained by integrating the transition probabilities between electron states in the conduction and valence bands over the states' energies:
\begin{multline}
\label{Material_gain}
g\left( F_{\bb n}, F_{\bb p} \right)=\frac{\pi e^2 \hbar }{n_{\bb{InAs}} m_0^2 \varepsilon_0}\frac{\left| M_{\bb b} \right|^2}{6\hbar \omega }
\int\limits_{-\infty }^{+\infty }{\rho_{\bb{cb}} (E) \rho_{\bb{vb}} (E-\hbar \omega)} \\
\left|M_{\bb{env}}(E,E-\hbar\omega) \right|^2 \left[ f_{\bb{cb}} ( E, F_{\bb n}) - f_{\bb{vb}} (E-\hbar \omega, F_{\bb p}) \right]dE.
\end{multline}
Here, $\rho_{\bb{cb}}(E)$ and $\rho_{\bb{vb}}(E)$ are the densities of states in the conduction and valence bands of the semiconductor, respectively, $f_{\bb{cb}}(E,F_{\bb n})$ and $f_{\bb{vb}}(E,F_{\bb p})$ are the Fermi-Dirac distribution functions for the conduction and valence bands, $M_{\bb b}$ is the average matrix element connecting Bloch states near the band edge~\cite{Coldren}, and $M_{\bb{env}}$ is the envelope matrix element calculated using Stern's model \cite{Casey-Stern}. In heavily doped semiconductors, $\rho_{\bb{cb}}$, $\rho_{\bb{vb}}$, $M_{\bb{env}}$ and consequently the material gain, strongly depend on the impurity concentration, while at low doping levels $M_{\bb{env}}$ is strongly peaked at the energy corresponding to the exact momentum conservation for the electron-photon system~\cite{CaseyPanish,Casey-Stern}, and equation~(10) is reduced to the that in the band-to-band transition model with the ${\bf k}$-selection rule.

\begin{acknowledgement}
This work was supported by the Russian Science Foundation (grant no. 14-19-01788).
\end{acknowledgement}



\end{document}